# How causal analysis can reveal autonomy in models of biological systems

**William Marshall[1], Hyunju Kim[2,3], Sara I. Walker[2,3] Giulio Tononi[1], Larissa Albantakis[1*]**

*[1†] Department of Psychiatry, University of Wisconsin, Madison WI*
*[2] BEYOND: Center for Fundamental Concepts in Science, Arizona State University, Tempe AZ*
*[3] School of Earth and Space Exploration, Arizona State University, Tempe AZ*

## Summary

Standard techniques for studying biological systems largely focus on their dynamical, or, more recently, their informational properties, usually taking either a reductionist or holistic perspective. Yet, studying only individual system elements or the dynamics of the system as a whole disregards the organisational structure of the system – whether there are subsets of elements with joint causes or effects, and whether the system is strongly integrated or composed of several loosely interacting components. Integrated information theory (IIT), offers a theoretical framework to (1) investigate the compositional cause-effect structure of a system, and to (2) identify causal borders of highly integrated elements comprising local maxima of intrinsic cause-effect power. Here we apply this comprehensive causal analysis to a Boolean network model of the fission yeast (*Schizosaccharomyces pombe*) cell-cycle. We demonstrate that this biological model features a non-trivial causal architecture, whose discovery may provide insights about the real cell cycle that could not be gained from holistic or reductionist approaches. We also show how some specific properties of this underlying causal architecture relate to the biological notion of autonomy. Ultimately, we suggest that analysing the causal organisation of a system, including key features like intrinsic control and stable causal borders, should prove relevant for distinguishing life from non-life, and thus could also illuminate the origin of life problem.

## Introduction

The emergence of life from non-living matter is widely regarded as one of the most challenging open problems in science, standing alongside other stubborn problems such as understanding agency, or characterising the nature of consciousness [1]. There are at present as many open questions as there are unique approaches to solving the origins of life. Most approaches tend to fall within three major categories: *historical*, dealing with how life first emerged on Earth; *synthetic*, addressing how to construct life from scratch; and *universal*, attempting to extract features such as 'aliveness' that might apply to life anywhere in the universe [2]. Progress in understanding the chemical bases of life, heralded by the molecular revolution of the 20$^{th}$ century, has set the focus of origins research on historical and synthetic approaches. Much less progress has been made with universal approaches. In part, this is due to a lack of alternative examples to contrast with the one chemical example of life we have (i.e., life on Earth): without even a second example it is difficult to separate potential universal features from those that are contingent features of life on Earth.

*Author for correspondence (albantakis@wisc.edu).
†Present address: Department of Psychiatry, University of Wisconsin, 6001 Research Park Blvd, Madison WI, 53719, USA

Despite intense debate on the matter, there have been numerous attempts to delineate a universal definition for life. Recent efforts have specifically focused on properties of life that might lay the ground for a theory of biology [3, 4, 5, 6]. Among the properties of life most amenable to this approach is *autonomy* [4, 5]. Roughly speaking, autonomous systems can be characterised as forming a unitary 'whole' from their own intrinsic perspective, and being able to maintain themselves in the face of changing internal and external states. That is, autonomous systems are integrated wholes with self-defined and self-maintained borders. This implies some notion of intrinsic *causal* control. It has been previously proposed that the origin of life might be identified as a transition in causal structure [6, 7]. If autonomy is universal to life, causation should be expected to play a prominent role in developing a more universal understanding of living processes: to be a living entity requires the power to regulate one's own internal state. Here we aim to bring quantitative rigor to the discussion by characterising the *intrinsic* causal mechanisms of a model of a biological system as a case study.

As a demonstrative example, we utilise a well-studied biological model: the Boolean network model of the fission yeast cell cycle [8]. Previous work has focused on the dynamics of this network, including the shape of the attractor landscape associated with cell-cycle function [8], network robustness and function [9], controllability for regulating function [10], and informational structure [11, 12]. While being a very simple model of biological function, the Boolean network model for the fission yeast cell cycle displays informational structure that is distinct from random networks with similar topological structure [11, 12], which suggests that it might be capturing some characteristic features of living systems. However, from the dynamics of the system alone we cannot understand why a particular state happens (i.e. what caused it), or how the system's components constrain each other (i.e. what caused what). A causal approach is necessary to know which elements or sets of elements form mechanisms with cause-effect power within the system [13, 14], and, thereby, to characterise architectures capable of intrinsic control.

Moreover, autonomy requires a system to construct its own 'umwelt' [15], causally separating itself from its environment. Nevertheless, living systems are open systems that dynamically and materially interact with their environment. For this reason, a purely dynamical approach cannot distinguish between a living entity and its environment. While classical information theory can be utilised to characterise an established individual within the framework of dynamical approaches [16], it does not inform how individuals can emerge in the first place. Only by employing counterfactual [17], interventionist [18] notions of causality can we properly decouple the intrinsic control a system has over itself from that of its environment and thereby hope to recover how such systems emerge.

To fully characterise the causal mechanisms involved in the fission yeast cell-cycle model we employ the formalism of integrated information theory (IIT) [19]. IIT provides a rigorous definition for intrinsic cause-effect power as integrated information (see Methods). By IIT's *composition* principle, any subset of elements within the system can be a mechanism of the system if its intrinsic cause-effect power is irreducible. Emergent mechanisms composed of more than one element are termed ‘high-order mechanisms’. Irreducibility is assessed by the integrated information φ (‘small phi’) of the subset of elements. If $\varphi > 0$, the subset in its current state constrains the past and future states of the system in an irreducible way (so that information is lost under any partition).

The set of all mechanisms and their constraints within the system comprises a system’s cause-effect structure. By IIT’s integration principle, a system has integrated information $\Phi > 0$ (‘big phi’) if its cause-effect structure is irreducible (so that mechanisms and constraints are lost under any system partition). A system with $\Phi > 0$ thus forms a unitary whole, as the elements within it constrain, and are being constrained by, the other elements of the system in an irreducible manner, above a background of external influences. While in principle many sets of elements within a larger system can have $\Phi > 0$, it is the local maxima of Φ that define where intrinsic causal borders emerge [20]. In this way, IIT provides a formal and quantitative framework to capture intrinsic properties of a system which may prove essential in distinguishing an entity from its environment [7, 20].

In the following, we demonstrate how the IIT analysis confirms established results regarding controllability and robustness of the Boolean network model of the fission yeast cell cycle, while providing a causal explanation for these properties. In addition, the analysis reveals previously overlooked attributes of the cell-cycle model, intrinsic control and causal borders, which are key features of biological autonomy [4, 5]. We propose that the IIT analysis has the capacity to provide a quantitative framework for establishing autonomy in biological systems, and outline the future work necessary to validate this proposal. Finally, we briefly discuss implications for the origins of life, and in particular, how the notions of integration and causal control might inform a universal definition of life that could be useful in classifying systems on the edge between life and non-life.

## Boolean network model of the fission yeast cell cycle

In modern cells, cellular division is tightly regulated and produces two daughter cells from a single parent cell through a cell cycle consisting of a sequence of four phases, G1 – S – G2 – M. During the phase G1, the cell grows, and if conditions are favourable, division begins. DNA is then replicated in the S stage. The G2 stage is a ‘gap’ between DNA replication (S phase) and mitosis (M phase) where the cell continues to grow. During the M stage, the cell undergoes mitosis, and two daughter cells are produced. The daughter cells then enter G1 again, thereby completing the full cycle.

The Boolean network model for the fission yeast cell-cycle process reproduces the protein expression states through each of these phases for nine proteins known to be key regulators governing the cell-cycle process [8]. Proteins are represented as nodes, and the links between two nodes are the inhibiting or activating biochemical interactions between pairs of proteins (mediated through regulation of gene expression). For each node, a binary state value ‘0’ or ‘1’ is assigned, which indicates the absence or presence of the particular protein, respectively. The successive states $S_i$ of a node *i*, are updated in discrete time steps by the following rule, which can be written as a function of the connectivity pattern, the states of nodes connected to *i* and the threshold $\theta_i$:

$$S_i(t+1) = \begin{cases} 1, & \Sigma_j a_{ij} S_j(t) > \theta_i \\ 0, & \Sigma_j a_{ij} S_j(t) < \theta_i \\ S_i(t), & \Sigma_j a_{ij} S_j(t) = \theta_i \end{cases} \quad (1)$$

where $a_{ij}$ denotes weight for a directed edge ( j → i ). For each edge, a weight is assigned according to the type of the interaction: $a_{ij} = -1$ for inhibition and $a_{ij} = +1$ for activation, and $a_{ij} = 0$ for no biochemical interaction. The threshold for all nodes in the network is 0 with the exception of $\theta_{Cdc2/13} = -0.5$ and $\theta_{Cdc2/13*} = 0.5$. If not stated otherwise, by '(cell-cycle) network' we always refer to the Boolean network model rather than the real biological system in the following.

The dynamical process of the network obtained by iterating the update rule in Eq. 1 reproduces the time sequence of protein expression states (shown in the Fig. 1B) corresponding to the four phases of the cellular division process. Here we define this particular progression of states as the biological sequence for the fission yeast cell-cycle model. Reproducing this sequence of states can be considered as the main function of the cell-cycle model. Previously, a subgraph of the network, called the 'backbone' motif, was discovered as the minimal set of connections necessary to exactly reproduce this biological sequence [9]. Other connections in the network, not included in the backbone, add robustness (associated with the shape of the attractor landscape, see below). Thus, for the fission yeast cell-cycle model, function is separable from robustness.

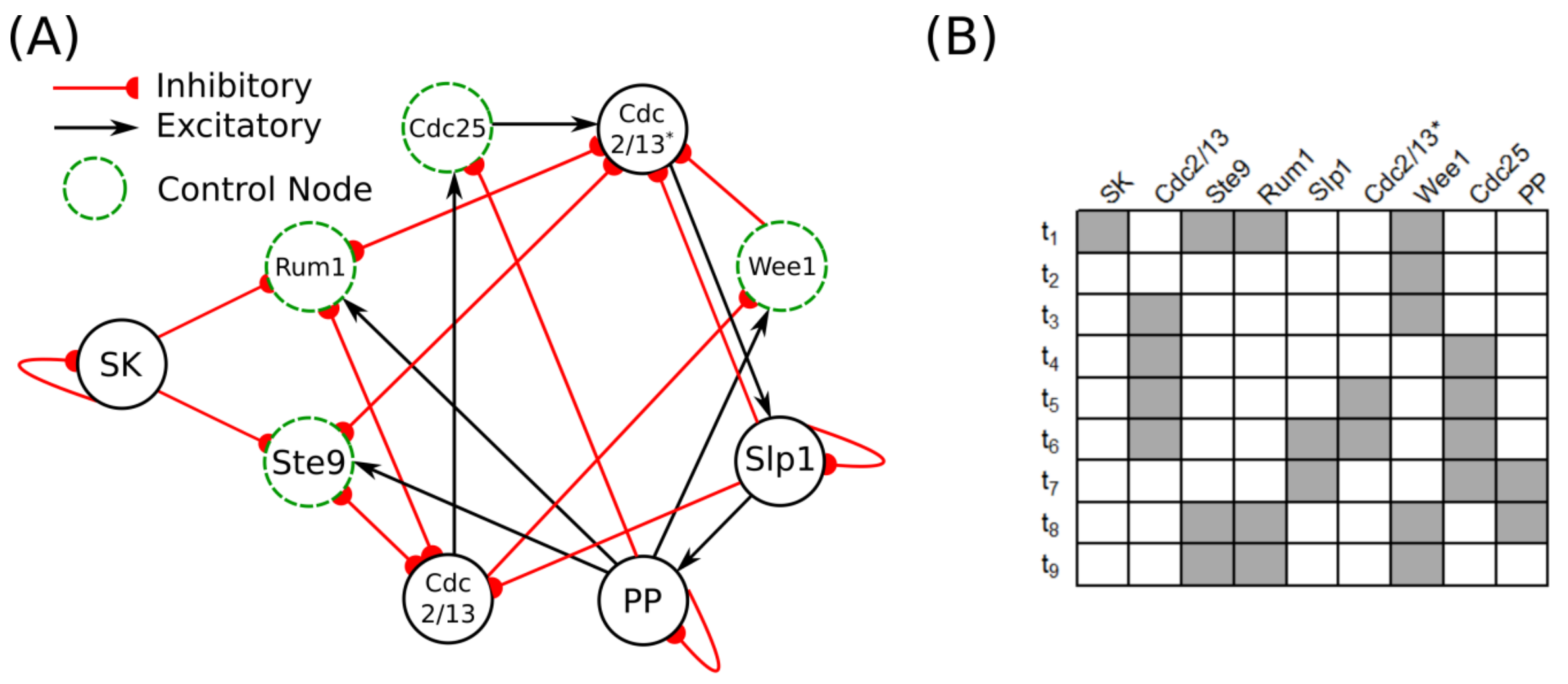

***Figure 1 – Fission yeast cell-cycle network.*** *(A) Network structure for a Boolean model of the fission yeast cell-cycle [6]. Control nodes of the system as identified in [8] are highlighted in green. (B) Sequence of nine states corresponding to the biological function of the network. The final state $t_9$ is the fixed point of the primary attractor in the system.*

The majority of studies on the fission yeast cell-cycle model are focused on its global dynamics in the state space of $2^9 = 512$ possible states for the nine-node network. The time evolution of the network initialised with each of the 512 possible states yields a flow diagram of network states that details *all* possible dynamical trajectories in the state space [8, 10]. In the flow diagram, each initial state converges into one of 13 possible attractors (12 fixed-point attractors and one limited-cycle attractor). The number of initial states that converges to each attractor is defined as its basin size and the attractor with the biggest basin size is called a primary attractor. For the network this corresponds to the final state of the biological sequence (herein called the biological attractor). The basin size of the biological attractor is associated with the robustness of the network, where networks with larger basins are more robust. Although this biological attractor is a fixed point in the state space of the model, in reality this state represents the completion of replication, at which point there are now two daughter cells each of which has returned to its initial state and the cycle can start again (here modelled as an `external' influence, by switching the SK node to ON).

The fission yeast cell-cycle model moreover features a subset of nodes {Ste9, Rum1, Wee1, Cdc25}—called the control kernel—that can regulate the network's robustness by external intervention [10]. The basin of the biological attractor naturally contains 378 of the 512 network states. To study the influence of individual elements and sets of elements on basin size, a 'pinning' operation is employed. External intervention is used to continuously pin the states of specific elements to their biological attractor state during network evolution to observe how the pinning influences the size of the attractor basin. The control kernel is the minimal set of nodes such that when pinned, the basin of the biological attractor is the entire state space of the network [10]**.** The presence of the control kernel also underlies the network's distinctive informational properties distinguishing it from random networks [11, 12]. Here, we further quantify the influence of individual nodes on the basin of the biological attractor by performing the pinning operation separately on each node in the network. We measured the change in the basin size of the biological attractor as a result of the time evolution with each individual node continuously pinned in its biological attractor state. This pinning operation performed on nodes Cdc2/13, Ste9 or Rum1 produces a larger basin size for the biological attractor, SK does not change the basin size, and all other nodes decrease the basin size as shown in the Fig. 2C.

In what follows, we apply the IIT formalism to perform a comprehensive causal analysis of the fission yeast cell-cycle model. We note that elementary interactions in the model (i.e., the elements' input-output functions and which elements can affect which other elements) have already been established through extensive prior experimental manipulation and observation [8]. The IIT analysis exposes the *compositional* causal structure hidden within the network of elementary interactions, by making its intrinsic, irreducible causal constraints explicit. Of course, the IIT analysis can only infer the causal structure of the specific model under consideration. The Boolean network model represents only a small fraction of the interactions that occur within a real fission yeast cell during the division process, and an extended model that includes these interactions would certainly reveal additional causal structure.

Applying the IIT analysis to the cell-cycle model reveals that the network has many high-order mechanisms and forms an integrated whole that is maintained through the phases of the cell cycle. The model's cause-effect structure elucidates how the system's high-order mechanisms drive regulation of the control kernel nodes and thus provides deeper understanding of how the network *internally* regulates its own function. We also demonstrate that the backbone motif of the cell-cycle network [9] is not similarly integrated through all phases. The function of the network can thus be detached from its robustness and integration.

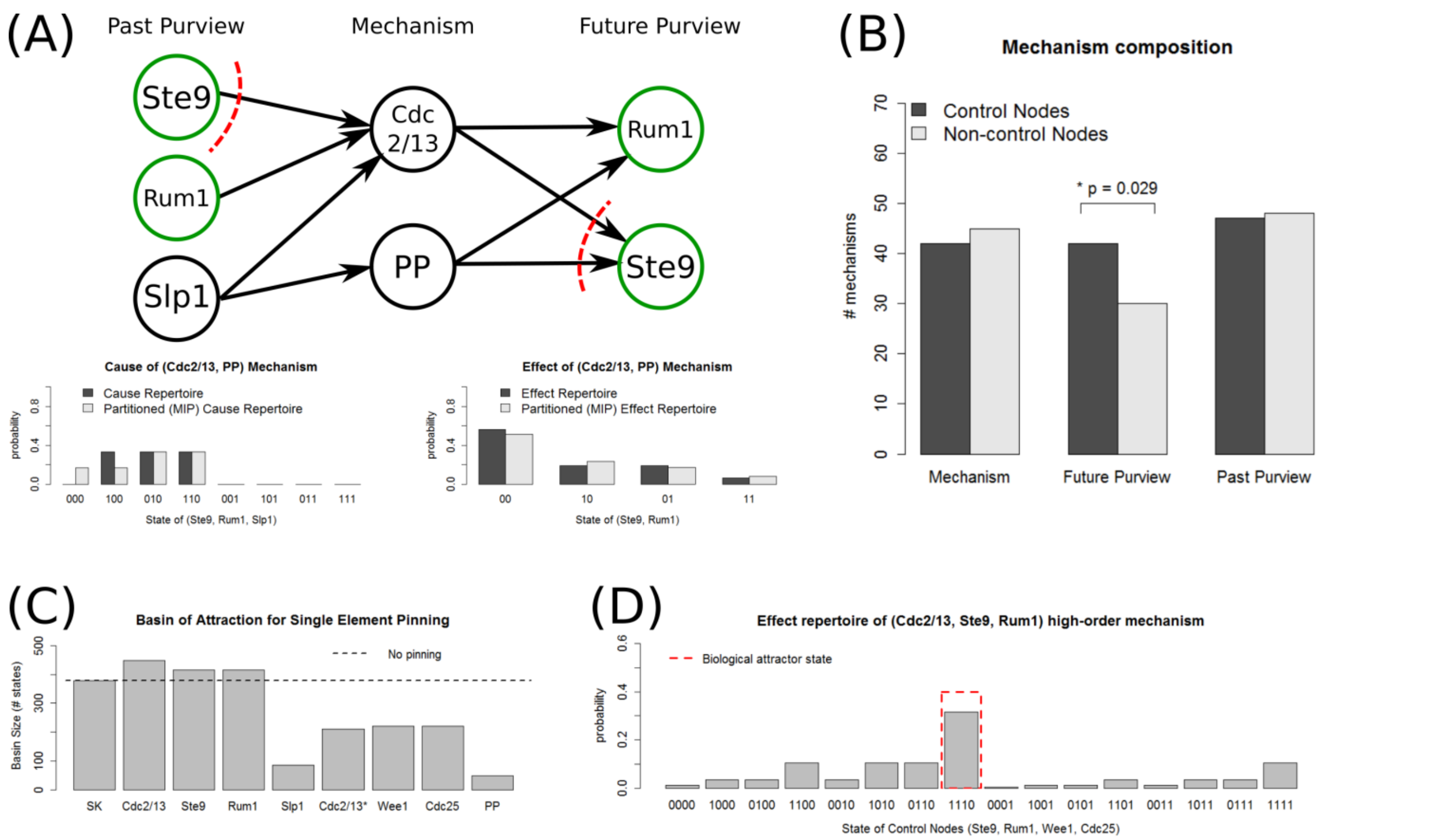

***Figure 2 – Composition of the cell-cycle network's cause-effect structure in the biological attractor state.*** *(A) Example high-order mechanism in the 8-node cell-cycle system. The two nodes Cdc 2/13 and PP together have joint, irreducible constraints on the past states of nodes Ste9, Rum1, and Slp1 and the future states of nodes Rum1 and Ste9, as indicated by their cause and effect repertoires. Since it is impossible to partition the mechanism and its purviews without losing part of the constraints, {Cdc2/13, PP} form a second-order mechanism in the system. Red dashed lines indicate the MIP corresponding to the partitioned cause and effect repertoires, as well as integrated information $\varphi = 0.15$. (B) Role of control nodes in the cause-effect structure. Despite their supposed functional role in the system, control nodes do not contribute to more system mechanisms than non-control nodes. Instead, control nodes appear in more future purviews, meaning that their future states are being constrained by more system mechanisms than those of the non-control nodes. (C) Contribution of individual nodes to the basin of the biological attractor as measured by the pinning operation (see text). (D) The third-order mechanism composed of the most influential elements from C, {Cdc2/13, Ste9, Rum1} constrains its future purview, consisting of the four control nodes, to be in state '1110' with highest probability, which corresponds to the biological attractor.*

# Results

The analysis of the cause-effect structure of the fission yeast cell-cycle model is broken down into three parts. First, we focus specifically on the biological attractor (fixed point $t_9$ in Fig. 1). We identify all mechanisms in the cause-effect structure and the specific way they constrain the system. Particular attention is paid to the role of extrinsically identified control nodes from the intrinsic perspective of the system. Next we identify local maxima of intrinsic cause-effect power $\Phi$ out of all the possible subsystems. Finally, we assess the robustness of these results across all nine states of the biological sequence. The element SK is an input, it signals to the rest of the network but never receives any feedback, and thus is not part of any integrated system. In all stages of the analysis, we focus on the remaining 8 nodes of the network.

## Cause-effect structure of the fission yeast cell-cycle model

The IIT causal analysis reveals that the cell-cycle network in its biological fixed point has 49 irreducible mechanisms ($\varphi > 0$), including all 8 possible first-order mechanisms and also 41 high-order mechanisms which irreducibly constrain the system's past and future states. A full list of mechanisms is given in Supplementary Table 1. An example of a second-order mechanism composed of Cdc2/13 and PP is shown in Fig. 2A. The mechanism irreducibly constrains the potential past states of Ste9, Rum1 and Slp1 (called its *past purview*) and the potential future states of Ste9 and Rum1 (called its *future purview*). The specific way that the mechanism constrains its past and future purviews is described by its cause and effect repertoires. The irreducibility of the mechanism is assessed by finding its minimum information partition (MIP, shown in red), and measuring the difference it makes to the repertoires. For this mechanism, we find that the integrated information is $\varphi = 0.15$. Overall, the first-order mechanisms have an average $\varphi$ value of 0.21, while the high-order mechanisms have an average $\varphi$ of 0.05.

Since the control kernel has been identified as a handle for extrinsic control [10], we further analyse the cause-effect structure of the cell-cycle network by investigating the contribution of control nodes to the mechanisms and their purviews (Fig. 2B). Of the 49 mechanisms, 42 include at least a single control node, and 45 contain at least a single non-control node. The average $\varphi$ for mechanisms with control nodes is 0.067, and with non-control nodes is 0.069. There is no significant difference between control nodes and non-control nodes, with respect to the composition of mechanisms or their $\varphi$. Furthermore, the set of all 4 control nodes is reducible ($\varphi = 0$) and thus does not form a high-order mechanism within the system. This means that the control nodes together do not have cause-effect power within the network. On the other hand, the control nodes are significantly over-represented in the future purviews of the mechanisms ($p = 0.029$ using nonparametric permutation test). Of the 49 mechanisms, 42 constrain the future state of at least one control node, while only 30 constrain the future state of at least one non-control node. The average $\varphi$ for a mechanism that constrains a control node is 0.071, while the average $\varphi$ for a mechanism that constrains a non-control node is 0.044.

In contrast to the control kernel, the three elements identified by the pinning procedure (Fig. 2C) do form an irreducible third-order mechanism within the system, meaning they have an irreducible effect within the system ($\varphi = 0.15$). Moreover, the specific way they constrain the potential future states of the system is to coerce the control nodes into the biological attractor (Fig. 2D). Given that setting the control kernel into a specific state ensures that the system enters its biological attractor, the result that intrinsic mechanisms constrain the control nodes into the biological attractor can be seen as a self-regulating property of the network.

## Local maxima of cause-effect power

Next we evaluate Φ of all possible subsystems of elements defined from the cell-cycle network in its biological fixed point. We are particularly interested in the local maxima of Φ, as these points correspond to subsystems (groupings of elements) where unique causal properties emerge. Evaluating all possible subsystems of the cell-cycle network, we identify 5 subsystems with $\Phi > 0$, three of which are local maxima of Φ (Fig. 3). The global maximum value of $\Phi = 0.431$ occurs for the whole system of 8 elements, this not only demonstrates the fission yeast cell-cycle network is an integrated whole, but that there is irreducible cause-effect power that can only be seen when the system is viewed as such. There are also local maxima for three nodes {Cdc2/13, Ste9, Rum1} with $\Phi = 0.090$, and for six nodes {Cdc2/13, Ste9, Rum1, Cdc2/13*, Wee1, Cdc25} with $\Phi = 0.001$, suggesting that these sets of elements form causally relevant subsystems within the cell-cycle network. Indeed, the system with $\Phi = 0.090$ corresponds exactly to the biologically relevant elements identified by the pinning procedure (Fig. 2C), suggesting that this subsystem plays an important role in the self-regulation of the cell-cycle network. Finally, the 6 node system with $\Phi = 0.001$ demonstrates the network is still integrated without PP and Slp1, albeit with greatly diminished cause-effect power.

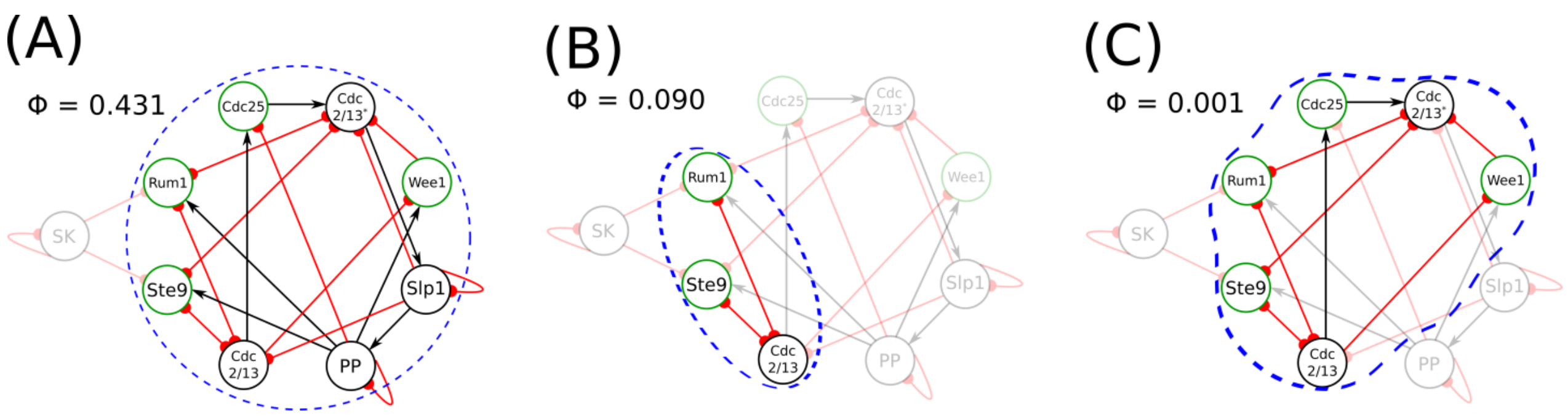

***Figure 3 – Local maxima of intrinsic cause-effect power in the biological attractor.*** *(A) The most irreducible cause-effect structure (global maximum) is specified by the largest strongly connected set of 8 elements, excluding only the input-node SK. (B) The three nodes that individually expand the biological attractor basin the most when set into their attractor state (Fig. 2C) also form an integrated system with many mutual constraints. (C) Taking SK, PP and Slp1 as background conditions yields another system with an integrated cause-effect structure comprised of six nodes.*

## Dynamics of causal properties

Cause-effect power quantifies how a system, by being in a particular state, constrains its potential past and future states through its intrinsic mechanisms. Thus, the causal properties of a system are state dependent: in different states, different mechanisms within the network may be irreducible, and they may constrain the past and future states of the system in different ways. To investigate the dynamical properties of cause-effect power, we analyse the cause-effect structure of the cell-cycle network in all the states comprised by its biological sequence. To contrast these results with standard functionalist approaches, we also analyse the backbone motif of the cell-cycle network [9], the minimal set of edges required to maintain biological function. Here, function is defined as reproducing the biological sequence of states associated with the real cell-cycle trajectory. There are 9 states in the biological sequence (Fig. 1B), however, the second state cannot be reached from within the 8-node system if SK is treated as a background condition. This means that this state has no causes from within the system (it is caused by the external element SK) and thus has undefined intrinsic cause-effect power. We thus restrict our analysis to the 8 well-defined states.

The dynamic analysis reveals the cell-cycle network to be robustly integrated. Over the course of the biological sequence, the full cell-cycle network of eight elements is integrated and, in fact, a local maximum of Φ in all 8 states (Fig. 4A). In each state, the cause-effect structure contains 8 first-order mechanisms, one for each element in the system. In addition, there are always high-order mechanisms in the cause-effect structure, as few as 10 for $t_1$ and as many as 111 at $t_4$ (Fig. 4B). Note, however, that the full network is not necessarily an integrated whole for states outside the biological sequence. For example, in the state where only SK is active, the system has only a single local maximum consisting of 6 elements {Cdc2/13, Ste9, Slp1, Cdc2/13*, Cdc25, PP}. The identified robust integration during the biological sequence is thus not simply the result of the model's network structure, but rather the specific state-dependent causal mechanisms that provide intrinsic control.

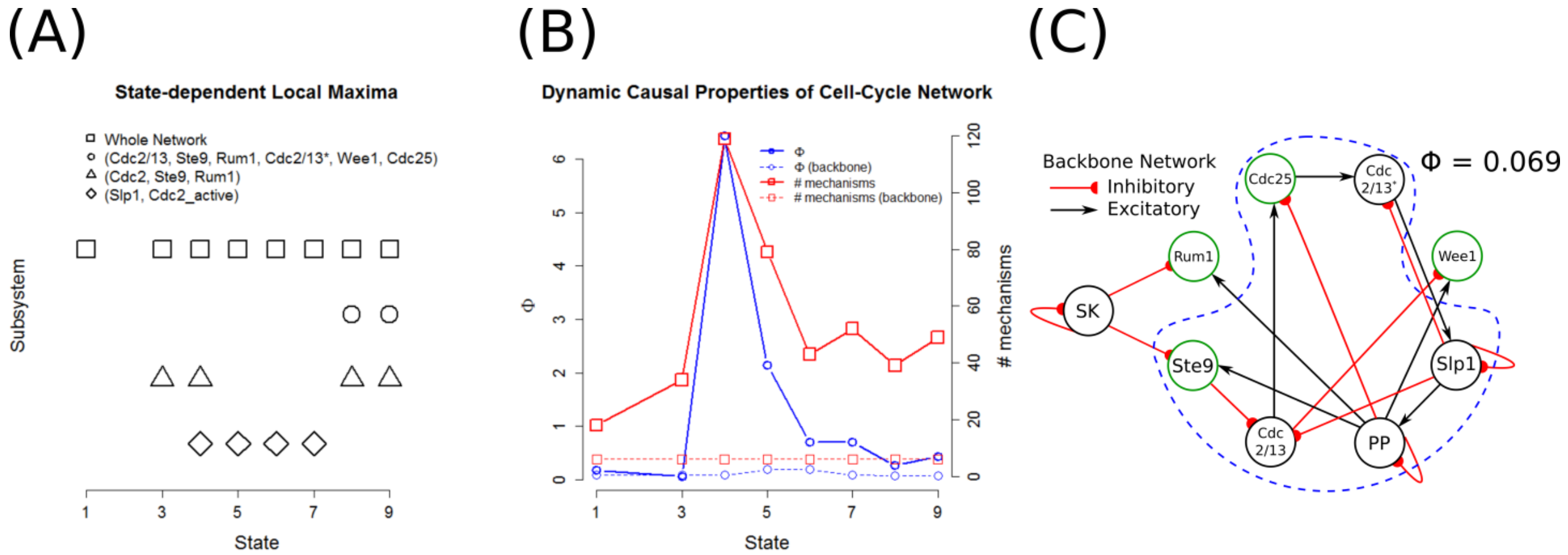

***Figure 4 – Stability of local maxima of Φ across biological cell-cycle network states.*** *(A) A total of four sets of elements form local maxima of Φ during the biological sequence. In all but one undefined state ($t_2$, see text) the 8 node system forms an integrated whole from the intrinsic perspective. (B) Comparison of Φ and number of mechanisms in the 8-node system and the largest strongly connected component of the cell-cycle network's backbone motif, which is sufficient to reproduce the function of the cell-cycle, but has a severely diminished cause-effect structure. (C) Backbone of the cell-cycle network [7].*

Moreover, in contrast to the full network, the backbone network (Fig. 4C) is not an integrated whole ($\Phi = 0$). The largest integrated subsystem of the backbone network contains 6 nodes (Rum1 and Wee1 are no longer integrated nodes in the reduced backbone network, as they lack outputs). This 6-node subsystem is only a local maximum of $\Phi$ in 3 of the 8 states. Furthermore, this system of 6 elements contains only 6 first-order mechanisms, and no high-order mechanisms in any of the biological states. This indicates that recapitulating the function of the biological cell cycle does not necessitate integration as observed in the full cell-cycle model. While both the full network and the reduced backbone motif network display the same function, only the full network is integrated, with consistent causal borders, for all states in the biological sequence. The identified stable borders in the biological sequence of the cell-cycle network thus emerge due to the intrinsic control that the system's mechanisms exert on each other, exhibiting self-maintenance, and not as a consequence of modelling the cell cycle's biological function.

# Discussion

In this work, we apply the causal framework of IIT to analyse the full cause-effect structure of the Boolean network model of the fission yeast cell cycle. We demonstrate how the framework's composition and integration principles identify emergent high-order mechanisms and emergent causal borders, defined as compositions of system elements that have irreducible cause-effect power, and as subsets of elements that define local maxima of intrinsic, irreducible cause-effect power, respectively. The results of the causal analysis reveal several properties of the cell-cycle network that cannot be uncovered by a purely dynamical approach, namely that the Boolean network model of the fission yeast cell cycle is a robustly integrated whole and is self-regulating. In summary, we have demonstrated how IIT's causal analysis can be applied to reveal defining features of biological autonomy – the ability of a physical system to self-define, and self-maintain its borders [4, 5].

Previous analysis of the cell-cycle network identified a control kernel that, when externally intervened upon, dictates the dynamics of the network [10]. However, studying the high-order mechanisms within the system reveals an intrinsic, causal mechanism which self-regulates not only the network's function, but also its casual borders, without the need for external manipulation. Implicit in this analysis is a rejection of the reductionist assumption according to which only individual micro elements have true cause-effect power. Reductionist approaches, as well as holist approaches that lack composition cannot account for how subsets of elements work together, constraining the system jointly and irreducibly to achieve a specific state transition [13, 19]. The study of high-order mechanisms may also help to inform genetic experiments on Schizosaccharomyces pombe cells, which so far have been primarily holist (sequencing the entire genome [21]), or reductionist (using 'gene knockout' to identify the functional role of individual genes [22]). Here, we find that the third-order mechanism composed of {CdC2/13, Spe9, Rum1} plays an important role for the self-regulation of the cell-cycle model (see Fig. 2D), and also that these three elements form a local maximum of cause-effect power (Fig. 3B). Each of these three genes have already been individually identified as important for cell-cycle function [23, 24], yet we predict that a multiple knockout of these three genes would have some effect on the function of the real cell cycle that is "greater than the sum of its parts".

Furthermore, a causal approach allows us to disentangle the intrinsic mechanisms within a system from the contributions of its external environment and thus establish self-defined borders necessary for autonomy. A functionalist approach, which only cares about what happens and not why it happens, has no way to distinguish the two contributions, and thus cannot firmly identify the borders between an entity and its environment. This distinction becomes important when comparing different networks with similar function. Comparing the causal structure of the cell-cycle network to that of its backbone motif reveals that only the former is robustly integrated. For the cell-cycle network, the whole network is a local maximum of $\Phi$ across all biological states. In contrast, the backbone network is not integrated ($\Phi = 0$), and the causal borders identified are not robust, varying across states. Thus, while the robustly integrated cell-cycle network can be viewed as a causally autonomous system, with self-defined and self-maintained borders, the backbone network lacks borders and the ability to maintain itself in the face of changing internal and external conditions. This establishes a distinction between function and integration, which may be important to the origins of life. A prebiotic system, such as an autocatalytic network or a protocell [25], even if functional, may not be integrated and thus would not be an autonomous system.

In general, the IIT analysis is computationally intense and requires extensive perturbational data, or an already established model of the basic system elements and their interactions. Suitable models of biological systems relevant for studying autonomy are still sparse. It is important to note that the cell-cycle network studied here is only a coarse model of a functional subpart of a Schizosaccharomyces pombe cell. While the cell-cycle model displays key features of biological autonomy, it lacks crucial components that are typically associated with the autonomy of the cell per se, for example, the cytoplasmic membrane. To confirm whether the set of elements comprising the cell-cycle network could still form an autonomous system once these additional components are taken into account, a properly extended model (supported by appropriate experiments) is required. One interesting question here is whether an extended model of an entire cell including the cytoplasmic membrane would reveal, as typically assumed, that the cytoplasmic membrane is within the self-defined borders of a cell. It might also turn out that the cytoplasmic membrane is not an intrinsic, causal part of the system, but rather provides advantageous background conditions to facilitate life.

The utility of the approach presented here is that it requires no *a priori* assumptions about what constitutes the borders of an entity: the model could be expanded to include any components within a supposed entity, as well as putative environmental variables, or even multiple entities. Once computationally feasible, and once a sufficiently detailed model of the cell is available, a demonstration that the IIT analysis indeed reveals appropriate causal borders in such an extensive model (i.e., which elements are in, and which are out), would provide strong evidence for the proposed approach. This quality, moreover, is a necessary prerequisite for any theoretical attempt to identify autonomy of a system on the edge between life and non-life. Causal control *internal* to a system's dynamics is another requirement for autonomy which has previously been proposed as a critical step in the emergence of living networks [26]. Indeed, it has been shown that control elements can emerge dynamically in catalytic networks [27]. Our analysis reveals that both internal control and causal borders can be identified using IIT. Since autonomy is likely a universal property of life, the kind of causal analysis presented here could place new constraints on models for the origin of life applicable to diverse chemical systems and help to identify at what point a living system has emerged.

# Methods

A system is a set of elements S in a state s, where each element has two or more states, inputs that influence its state and outputs which influence the state of system elements. A system has a corresponding transition probability function p which describes the probabilities with which the system transitions from one state to another for all possible system states. The transition probability function provides the basis for our causal analysis. For a given system, the transition probability function has to be determined by experiments involving intervention and manipulation of variables [18], which either identify the input-output relationship of the system's elements, from which the transition probability function can be constructed, or directly measure the state-to-state transition probabilities of the system. A properly determined transition probability function should satisfy the Markov property. In addition, we require that the current states of elements are independent, conditional on the past state of the system (prohibiting instantaneous, non-causal interactions). For any two subsets of S, called the mechanism Y and the purview Z, we can define the cause and effect repertoires of Y over Z, that is, how Y in its current state $y_t$, constrains the potential past or future states of Z,

$$p_{\text{cause}}(z_{t-1}|y_t) = \frac{1}{K_{\text{cause}}} \prod_i \frac{\sum_{z^c} p\left(y_{i,t} \middle| do(z_{t-1}), do(z^c)\right)}{\sum_s p\left(y_{i,t} \middle| do(s)\right)}$$

and

$$p_{\text{effect}}(z_{t+1}\,|y_t) = \frac{1}{K_{\text{effect}}} \prod_i \sum_{y^c} p\left(z_{i,t+1} \middle| do(y_t), do(y^c)\right).$$

where $K_{\text{cause}}$ and $K_{\text{effect}}$ are normalising constants [18, 19, 28]. The superscript 'c' labels the complement of a set of elements, e.g. $Z^c = S \backslash Z$.

The integrated cause-effect information of Y is then defined as the distance between the cause-effect repertoires of the mechanism, and the cause-effect repertoires of their minimum information partition (MIP) over the purview that is maximally irreducible,

$$\varphi_{\text{cause}} = \max_Z \min_{cut} D\left(p_{\text{cause}}(z_{t-1}|y_t), p_{\text{cause}}^{cut}(z_{t-1}|y_t)\right),$$

$$\varphi_{\text{effect}} = \max_Z \min_{cut} D\left(p_{\text{effect}}(z_{t+1}|y_t), p_{\text{effect}}^{cut}(z_{t+1}|y_t)\right),$$

where cut is a partition of the mechanism into two parts, and $p^{\text{cut}}$ the repertoire under the partition,

$$cut = \{Y_1, Z_1, Y_2, Z_2\},$$

$$p^{cut}(z|y) = p(z_1|y_1) \otimes p(z_2|y_2).$$

The integrated information of the mechanism is the minimum of its corresponding integrated cause and effect information,

$$\varphi = \varphi(y) = min(\varphi_{\text{cause}}, \varphi_{\text{effect}}).$$

All sets of elements with $\varphi > 0$ are irreducible mechanisms M within the system and thus contribute to the system's cause-effect structure $C = \{\{p_{\text{cause}}, p_{\text{effect}}, \varphi\}_M\}$, the set of all irreducible cause-effect repertoires with their corresponding $\varphi$ values. The integrated information of the entire system is then defined as the distance between the cause-effect structure of the system, and cause-effect structure defined by its minimum information partition (MIP), eliminating constraints from one part of the system to the rest:

$$\Phi = \min_{cut} D(C, C^{cut})$$

For both the integrated information of a mechanism (φ) and the integrated information of a system (Φ), the distance metric used is the earth mover's distance [19, 29]. Finally, if S is a subset of elements within a larger system, all elements outside of S are considered as part of the environment and are conditioned on their current state throughout the causal analysis. All computations for this study were performed by the PyPhi software package [30], using the "CUT_ONE_APPROXIMATION" to **Φ.**

# Additional Information

**Authors' Contributions**
WM, GT and LA conceived of the experiment. WM and HK performed the analyses. WM, SIW and LA interpreted the results. All authors contributed to drafting and critically editing the manuscript.

**Competing Interests**
SIW is the guest editor for the "ORIGINS" special issue. This manuscript was given to an alternate editor for assessment.

**Funding Statement**
WM, LA and GT are supported by the Templeton World Charities Foundation (Grant #TWCF 0067/AB41).